\documentclass{Interspeech}

\usepackage{cite}
\usepackage{amsmath,amssymb,amsfonts}
\usepackage{graphicx}
\usepackage{textcomp}
\usepackage{xcolor}
\usepackage{hyperref}
\usepackage{url}
\usepackage{todonotes}
\usepackage[utf8]{inputenc}
\usepackage[T1]{fontenc}
\usepackage[english]{babel}
\usepackage{subcaption}
\usepackage{changes}
\usepackage{algorithm}
\usepackage{algpseudocode}
\usepackage{multirow}
\usepackage{tabularx}
\usepackage{booktabs}

\def\NemoIPL{TopIPL}

\interspeechcameraready

\title{ Unified Semi-Supervised Pipeline for Automatic Speech Recognition}

\author[affiliation={1}]{Nune}{Tadevosyan}
\author[affiliation={1}]{Nikolay}{Karpov}
\author[affiliation={1}]{Andrei}{Andrusenko}
\author[affiliation={2}]{Vitaly}{Lavrukhin}
\author[affiliation={2}]{Ante}{Jukic}

\affiliation{}{NVIDIA}{Armenia}
\affiliation{}{NVIDIA}{USA}
\email{ntadevosyan@nvidia.com}

\keywords{Speech recognition, Pseudo-labeling, Portuguese, Spanish, Armenian}

\usepackage{comment}

\begin{document}

\maketitle

\begin{abstract}

Automatic Speech Recognition has been a longstanding research area, with substantial efforts dedicated to integrating semi-supervised learning due to the scarcity of labeled datasets. However, most prior work has focused on improving learning algorithms using existing datasets, without providing a complete public framework for large-scale semi-supervised training across new datasets or languages. In this work, we introduce a fully open-source semi-supervised training framework encompassing the entire pipeline: from unlabeled data collection to pseudo-labeling and model training. Our approach enables scalable dataset creation for any language using publicly available speech data under Creative Commons licenses. We also propose a novel pseudo-labeling algorithm, TopIPL, and evaluate it in both low-resource (Portuguese, Armenian) and high-resource (Spanish) settings. Notably, TopIPL achieves relative WER improvements of 18–40\% for Portuguese, 5–16\% for Armenian, and 2–8\% for Spanish.
\end{abstract}

\section{Introduction}

Training of Automatic Speech Recognition (ASR) models requires large amounts of labeled data, which is often expensive and time-consuming to obtain. Semi-supervised learning mitigates this challenge by leveraging large-scale unlabeled speech data. However, acquiring such data presents its own difficulties, since data collected from real-world recordings is often noisy and may contain speech in multiple languages, background music, or other disruptions. Online platforms like YouTube offer a rich and diverse repository of speech recordings that, when properly utilized, can significantly enhance model performance. 

Early works in semi-supervised ASR \cite{synnaeve2020endtoend, kahn2020self} introduced simple pseudo-labeling algorithms, where a model is first trained on labeled data and then used to generate pseudo-labels (PLs) for unlabeled data, which are subsequently incorporated into training. 
More recent approaches have evolved into iterative strategies \cite{xu20b_interspeech, berrebbi2023continuous, likhomanenko21b_interspeech}, refining pseudo-label quality over multiple training cycles. To further enhance these algorithms, teacher-student frameworks have been widely adopted, where the teacher model aggregates student weights to improve accuracy. Algorithms such as exponential moving average (EMA) and momentum pseudo-labeling (MPL) \cite{higuchi21_interspeech, manohar2021kaizen, higuchi2022momentum, higuchi2022advancing} enhance pseudo-label stability, while InterMPL \cite{higuchi2023intermpl} introduces an additional intermediate loss.  Extensive research has also been conducted on pseudo-label filtering to improve reliability, including fusion scoring \cite{park20d_interspeech}, Levenshtein-based filtering \cite{zhang22u_interspeech}, and confidence-based error detection \cite{zhu2023alternative}. Additionally, uncertainty estimation techniques such as dropout-based self-training \cite{khurana2021unsupervised} and gradient masking \cite{ling2022improving} have been explored to stabilize training, reducing overfitting risks and ensuring more robust learning.

Despite these advancements, several shortcomings remain. Firstly, many state-of-the-art approaches are not open-source,  limiting reproducibility and slowing research progress. Secondly, most works rely on controlled, well-curated datasets rather than real-world speech data, making it difficult to assess these methods perform in practical scenarios with noisy and diverse audio. As a result, their applicability to real-world ASR remains uncertain. Finally, prior work often tackles isolated aspects of the data pipeline, like pseudo-label generation or filtering, without offering a complete system that researchers can easily replicate and extend, complicating adoption.

In this work, we aim to bridge the gap between research-driven data curation and real-world ASR applications. We introduce a fully open-source and reproducible pipeline that efficiently automates large-scale speech data collection from social-media platforms and similar publicly available sources, significantly reducing manual filtering and improving accessibility for low-resource languages. Furthermore, we propose \NemoIPL{}.\footnote{\href{https://github.com/nune-tadevosyan/NeMo/tree/TopIPL}{\url{https://github.com/nune-tadevosyan/NeMo/tree/TopIPL}}}, a novel and effective pseudo-labeling algorithm that integrates a dynamic cache with a student-teacher technique. Unlike other iterative approaches, \NemoIPL{} generates pseudo-labels only after the model has fully converged on labeled data, ensuring consistently high transcription accuracy while minimizing the risk of divergence during training. To maintain training consistency, our method updates the cache at each epoch’s end with a given probability and leverages the student’s top-$N$ checkpoints, selected by validation performance, to retain the most informative model states.

Notably, \NemoIPL{} achieves significant improvements over a baseline trained only on labeled data, without requiring explicit filtering or uncertainty estimation, demonstrating strong robustness to real-world noisy data. We validate our approach in both low-resource (Portuguese) and high-resource (Spanish) settings using public data under the Creative Commons license, highlighting the effectiveness of pseudo-label training even when unlabeled data is out-of-domain. Evaluations on the LibriSpeech-100/LibriSpeech-860 setup further show that \NemoIPL{} achieves up to  25\% reduction in error rates, outperforming previous works and reinforcing its efficacy in semi-supervised ASR. Main contributions of this study are:
 
\begin{itemize}
    \item Automated preprocessing of large-scale speech datasets sourced from open online platforms.

    \item TopIPL, a novel pseudo-labeling algorithm that delivers significant accuracy improvements across all experimental setups.

    \item Open-source release of the data collection pipeline and \NemoIPL{}, promoting reproducibility and accessibility.

\end{itemize}

\begin{figure}[t]
    \centering
    \begin{minipage}{0.45\textwidth}
        \centering
        \includegraphics[width=\linewidth]{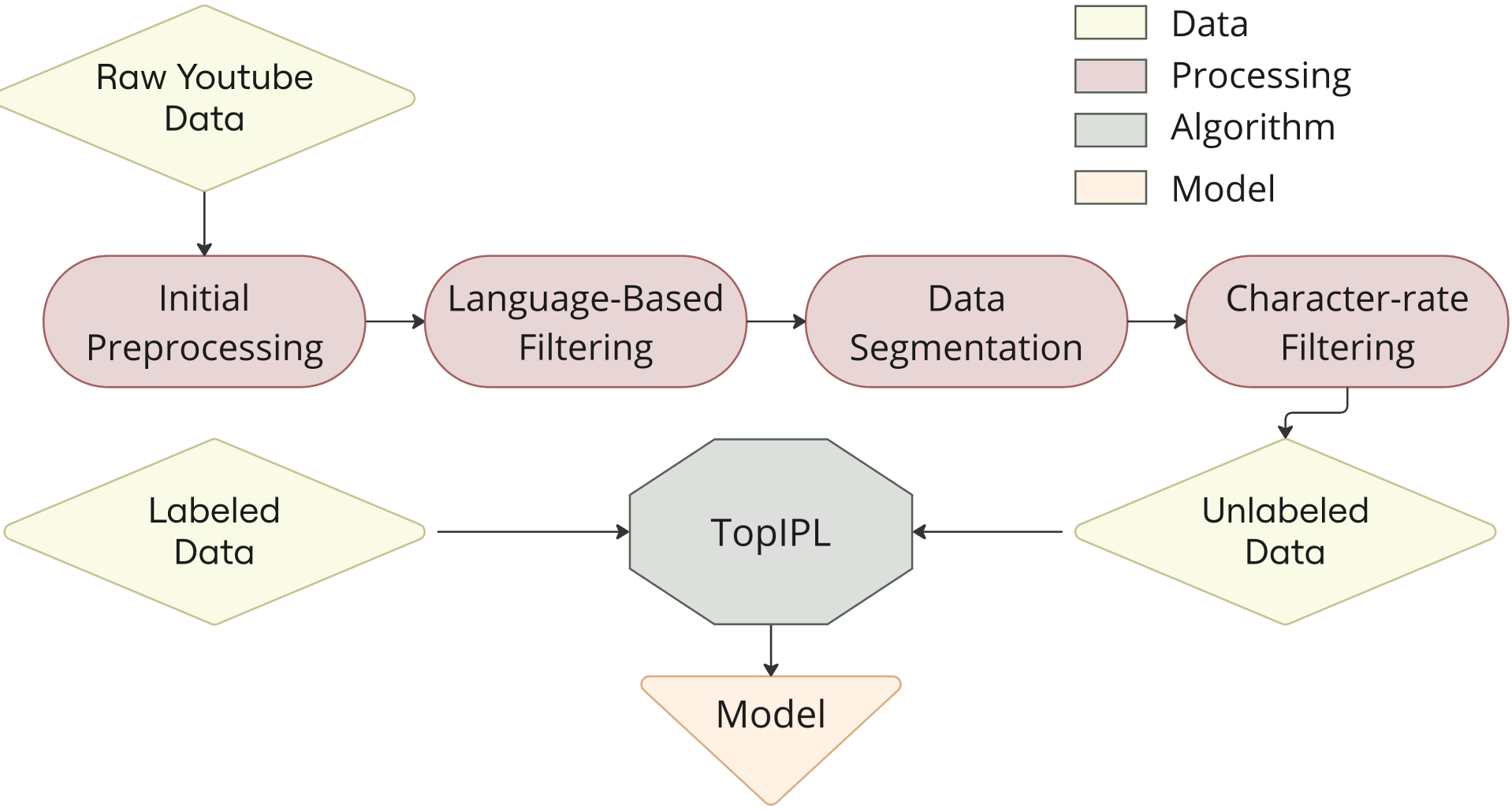}
        \caption{Proposed semi-supervised pipeline for training  ASR models.}
        \label{fig:Workflow}
    \end{minipage}
\end{figure}

\section{Method} 
We follow a structured pipeline\footnote{\href{https://github.com/NVIDIA/NeMo-speech-data-processor/tree/main/dataset_configs/portuguese/unlabeled}{\url{https://github.com/NVIDIA/NeMo-speech-data-processor/tree/main/dataset_configs/portuguese/unlabeled}}} to process large-scale speech data, ensuring clean and language-specific training samples. Our approach can be applicable to any language and integrates NeMo tools, including the Speech Data Processor (SDP)\footnote{\url{https://github.com/NVIDIA/NeMo-speech-data-processor}}, to automate data collection and preprocessing. The main stages of the proposed pipeline are illustrated in Figure~\ref{fig:Workflow}.

\subsection{Initial preprocessing}
To leverage large-scale unlabeled speech data, we source a subset of recordings from Yodas \cite{li2024yodasyoutubeorienteddatasetaudio} and YouTube Commons \cite{PleIAs_YouTubeCommons} datasets shared under the Creative Commons\cite{CCBY4.0} license. The preprocessing stage begins by converting Opus audio files into a standardized format using single-channel 16\,kHz WAV files. To manage variability in audio duration and reduce computational overhead of handling long audio, we discard long recordings exceeding 4,000 seconds.

\subsection{Language-based filtering}
Online recordings often contain speech in multiple languages, requiring filtering to retain only the target language. After formatting, we perform language identification using \mbox{AmberNet}~\cite{jia2022ambernet}\cite{nvidia_langid_ambernet} model, which runs on CUDA and employs a segment-based voting mechanism. This step ensures that only recordings in the intended language are included in the dataset. 

\subsection{Data segmentation}
Audio segmentation is typically performed using voice activity detection algorithms, which segment speech based on pauses between active segments. To enhance segmentation accuracy, we further employ NeMo Forced Aligner \cite{nemo_forced_aligner}, which generates word-level timestamps by aligning predicted transcriptions (with punctuation and capitalization) from a Fast Conformer Hybrid Transducer-CTC model. Sentences are reconstructed using punctuation as boundary markers, and segments ranging from 1 to 20 seconds are extracted. Each audio file is then cropped accordingly to produce clean, structured training data. 
\begin{algorithm}
\resizebox{0.95\columnwidth}{!}{%
\begin{minipage}{\columnwidth}
\caption{TopIPL Pseudo-Labeling Algorithm}
\label{alg:enhanced_pl}
\textbf{Data:} Labeled data $\mathcal{L} = \{x_i, y_i\}$, Unlabeled data $\mathcal{U} = \{x_j\}$ \\
\textbf{Result:} Student model $S$, Teacher model $T$

\begin{algorithmic}[1]

    \State \textbf{For} $n_\text{epochs}$ \textbf{do:}
    \State \quad Train $S$ on $\mathcal{L}$ with data augmentation until full convergence.
    
    \State \textbf{At the end of} $n_\text{epochs}$ \textbf{:}
    \State \quad{Generate} pseudo labels $\{\hat{y}_j\}$ for the cache $\mathcal{C}$ using $S$.
    \State \quad Store all $\mathcal{P} = \{x_j, \hat{y}_j\}$ data in the cache $\mathcal{C}$.
    
    \State \textbf{For} $m_\text{epochs}$ \textbf{do:}
    \State \quad Train $S$ on $\mathcal{L}$ and $\mathcal{P}$ with data augmentation.
    \State \quad Update $\mathcal{C}$ from $\mathcal{U}$ with probability $p_\text{cache}$ using $S$.
    
  \State \textbf{At the end of} $m_\text{epochs}$ \textbf{:}
    \State \quad Initialize $T$ with the current weights of $S$.
    
    \State \textbf{For} $m_\text{epochs}$ \textbf{do:}
    \State \quad Train $S$ on $\mathcal{L}$ and $\mathcal{P}$ with data augmentation.
    \State \quad Update $T$ by averaging the weights of 
    \newline     top-$N$ checkpoints of $S$.
    \State \quad With probability $p_\text{cache}$, update $\mathcal{C}$ with $\mathcal{P}$ generated by $T$.
\end{algorithmic}
\end{minipage}%
}
\end{algorithm}
\subsection{Character rate-based filtering} \label{subse:cr_filtering} 
Given the nature of crowdsourced data, many recordings contain noise, background speech, or music. Initially, we retain all audio files to assess their impact on training. However, additional filtering may be applied using baseline ASR models. A baseline model is used to generate a preliminary transcription and the corresponding character rate (CR), the number of characters per second, is computed. Audio files with exceptionally low CRs (below 5) or high CRs (above 21) are filtered out to remove the extreme cases. Similarly, filtering can be performed based on WER using a predefined threshold to remove low-quality transcriptions.

\subsection{TopIPL}
With a structured dataset in place, we transition to the pseudo-labeling stage using \NemoIPL{}. In previous approaches such as \cite{likhomanenko21b_interspeech}, pseudo-labels were generated gradually during training on labeled data $\mathcal{L}$, evolving as the model learned. However, in our method, the baseline model is first trained exclusively on $\mathcal{L}$ until full convergence, ensuring a strong foundation. Only after this point are pseudo-labels generated for the entire unlabeled dataset $\mathcal{U}$ in a single pass. This guarantees high-quality transcriptions, minimizing divergence risk from noisy labels.

To maintain training consistency, pseudo-labels are stored in a dynamic cache, which is updated as training progresses. As the model continues to improve, pseudo-labels are refreshed with probability $p_\text{cache}$, ensuring that training benefits from more accurate transcriptions over time. Through extensive experimentation, we determined that $p_\text{cache}=0.2$ yields the best results. The model improves so significantly after each epoch that there is no need to retain the best checkpoints, since each successive checkpoint surpasses the performance of the previous one. However, after several epochs, WER improvements plateau on validation sets, signaling a diminishing return from further training in this stage.

At this point, training transitions into a teacher-student phase. In this phase, we track and retain the top-$N$ checkpoints that achieve the best validation performance from the continuously improving model. Unlike conventional methods \cite{higuchi21_interspeech,manohar2021kaizen} that update the teacher model using moving averages, we take a more selective approach by updating the teacher only with the averaged top-$N$ student checkpoints. This ensures that the most informative model states are captured, leading to more stable and higher-quality pseudo-labels. As a result, the teacher model continuously provides refined supervision, further enhancing the effectiveness of \NemoIPL{}.

\section{Experimental Setup}

For our experiments, we utilize the NVIDIA Fast Conformer encoder \cite{rekesh2023fast}, which is 2.8 times faster than the standard Conformer encoder \cite{gulati20_interspeech}. To accelerate model convergence, we employ two decoding heads: Connectionist Temporal Classification (CTC) \cite{graves2006connectionist} and Recurrent Neural Network Transducer (RNNT) \cite{rao2017exploring}, with the model's loss function being a weighted average of both\cite{noroozi2024stateful}. The model consists of approximately 114 million parameters, and this remains consistent across all training setups. Pseudo-labels are generated using the CTC head due to its speed, but performance improvements are seen for the RNNT head as well. We set the dropout rate to 0.1 and applied spectrogram augmentation\cite{park19e_interspeech} to both labeled and unlabeled data with frequency mask of 2 and time mask of 10. 

\subsection{Training details}

\subsubsection{Spanish language}
Training setup for the Spanish model utilizes four labeled datasets(1400h): Mozilla Common Voice (MCV) \cite{ardila2019common}, Multilingual LibriSpeech (MLS) \cite{pratap20_interspeech}, Fisher, and VoxPopuli (VP) \cite{wang2021voxpopuli}. To leverage prior knowledge, we initialize the encoder weights with a pre-trained English model \cite{NVIDIAFastConformer}. Baseline training is performed over $n_\text{epochs} = 300$ epochs using only the labeled datasets. In \NemoIPL{} experiments, we extend the training with an additional 10k hours of unlabeled data, refining the model over $m_\text{epochs} = 50$ epochs across various setups.

\subsubsection{Portuguese language}
Unlike Spanish, our Portuguese model relies primarily on MCV (80h) as the labeled dataset. Due to incomplete transcriptions in MLS(160h), we treat it as an unlabeled source rather than a supervised dataset. We initialize the Portuguese encoder with the pre-trained Spanish model’s encoder. The model is trained for $n_\text{epochs} =  50$ epochs, which is sufficient to reach stable convergence. In the \NemoIPL{} experiments, we extend the training with an additional 2.5k hours of unlabeled data, refining the model over $m_\text{epochs} = 50$ epochs across various configurations. The  mTEDx\cite{salesky2021multilingual} and CORAA\cite{junior2021coraa} datasets are reserved for evaluation only as they are licensed under non-commercial \mbox{CC BY-NC-ND 4.0} license.

\subsubsection{Armenian language}
Our training setup for the Armenian model utilizes two labeled datasets: MCV (50h) and Fleurs (7h)\cite{Conneau2022FLEURS}. As for Spanish, we utilize encoder weights of a pre-trained English model \cite{NVIDIAFastConformer}. Baseline training is performed over $n_\text{epochs} = 200$ using 57h of labeled speech. As Armenian is a low-resource language we could obtain only 145h of unlabeled data and \NemoIPL{} training is conducted over $m_\text{epochs} = 50$.

\begin{figure}[t]
    \centering
    \includegraphics[width=\linewidth]{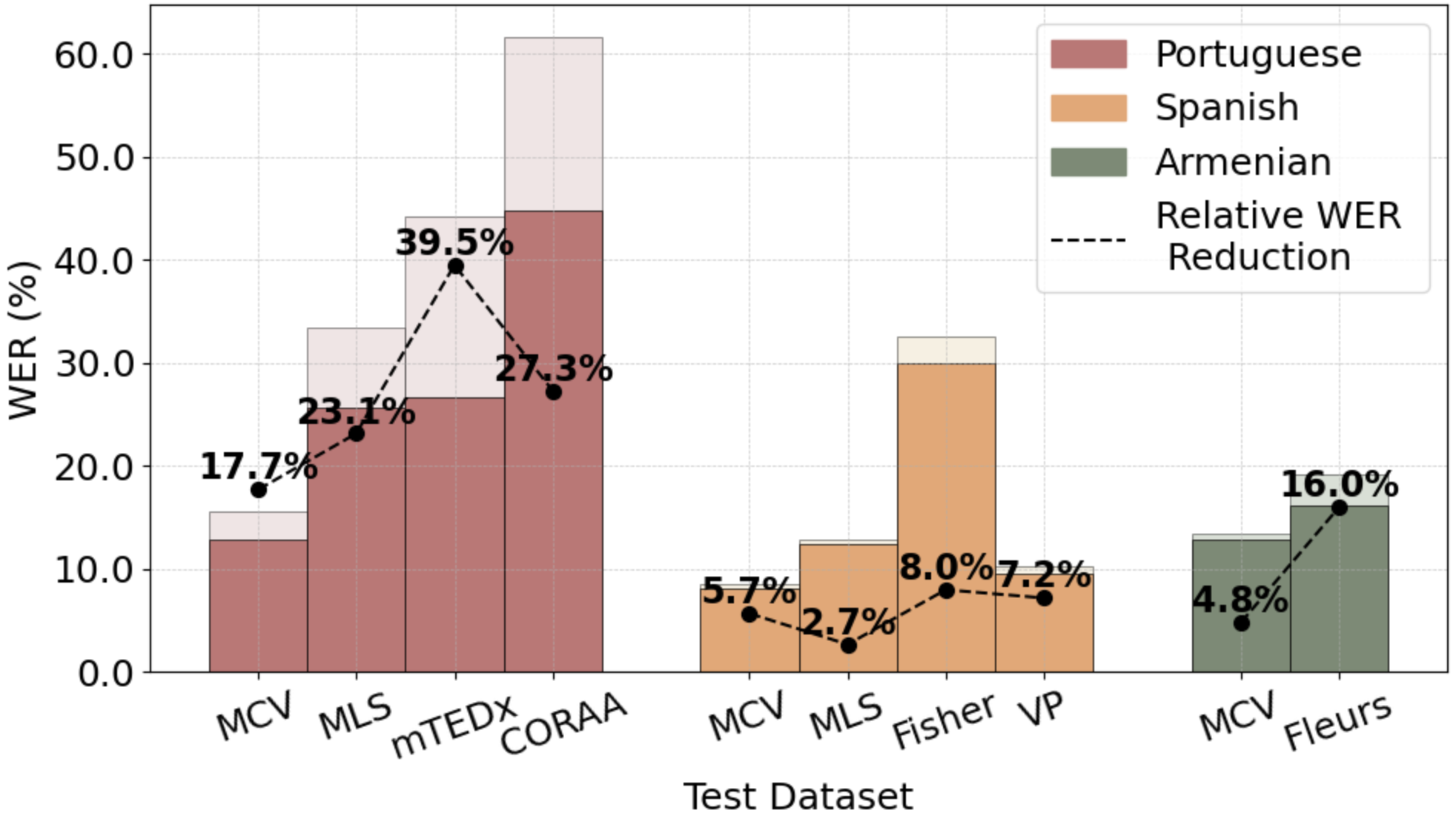}
    \caption{WER improvements across languages and datasets. The transparent part represents the baseline performance, while the opaque part is the performance after applying \mbox{TopIPL}.}
    \label{fig:graph}
\end{figure}

\begin{table}[ht]
    \centering
    \caption{Performance evaluation of models trained on Spanish data in terms of WER (\%). Bold values denote the best result in each group. }
    \label{tab:results_spanish}
    \vspace{-0.2cm} 
    \resizebox{\columnwidth}{!}{
    \begin{tabular}{l@{\hskip 3pt}|c@{\hskip 3pt}|c@{\hskip 3pt}c@{\hskip 3pt}c@{\hskip 3pt}c}
        \toprule
        \textbf{Model} & \textbf{Unlabeled} & \textbf{MCV} & \textbf{VP} & \textbf{MLS} & \textbf{Fisher} \\ 
        \midrule
        Baseline CTC & \textbf{-} & 8.61 & 10.25 & 12.8 & 32.55 \\
        Baseline RNNT & \textbf{-} & 7.79 & 10 & 12.6 & 31.72 \\
        \midrule
        \multirow{5}{*}{\shortstack[l]{First stage\\ No CR Filtering}} 
         & \textbf{2k h} & \textbf{8.29} & 9.51  & \textbf{12.5} & 30.39 \\
         & 4k h & 8.52 & 9.55 & 12.73 & 30.52  \\
         & 6k h & 8.43 & \textbf{9.47} & 12.98 & 29.84 \\
         & 8k h & 8.56 & 9.59 & 13.01 & 29.63 \\
         & 10k h & 8.81  & 9.66 & 12.82 & \textbf{29.49 } \\
         \midrule
        \multirow{2}{*}{+ CR Filtering} & 2k h & 8.24  & 9.67 & \textbf{12.38} & 30.47 \\
         & 8k h & 8.69  & 9.54& 12.82  &  29.88 \\
        \midrule
     \multirow{1}{*}{TopIPL CTC} & 2k h & \textbf{8.12} & \textbf{9.51} & \textbf{12.45} & \textbf{29.92}\\
        
        TopIPL RNNT &  2k h & \textbf{6.86} & \textbf{8.54} & \textbf{11.66} & \textbf{27.91} \\
         \midrule
    \end{tabular}%
    }
\end{table}

For all experiments, we utilized the Adam optimizer and a cosine learning rate scheduler. Hyperparameters were tuned based on validation performance across all datasets using WER. We experimented with maximum learning rates ranging from $5 \times 10^{-3}$ to $1 \times 10^{-3}$ and found that, at each stage, the best setup would be if maximum learning rate was half of the previous stage.

\section{Results}
In Figure~\ref{fig:graph}, we present the performance of models trained using the \NemoIPL{} algorithm across three different languages. The most significant improvement is observed in Portuguese, where relative WER reduction reaches up to 40\%. Here, labeled data is scarce, but a sufficient amount of unlabeled data is available, allowing the model to leverage self-supervised learning effectively. Armenian also demonstrates substantial gains, achieving a 16\% relative WER improvement despite having a more limited pool of unlabeled data (145h). We believe that with access to more unlabeled data, these improvements could be further amplified.   Lastly, the Spanish model, which was already robust due to a well-balanced labeled dataset, still benefits from our approach, showing an 8\% relative WER improvement. All evaluations were conducted without incorporating any external language model.

\subsection{Spanish language}
Table~\ref{tab:results_spanish} presents the performance evaluation of Spanish models trained with different amounts of unlabeled data at various stages of the algorithm. Our analysis show that adding 2k hours of unlabeled data improves WER across all datasets . However, adding more data beyond this point yielded less predictable benefits, with some datasets (e.g., MLS) showing signs of performance degradation. Likely this is because the labeled data alone is enough to create a strong model. Alternatively, the decline in performance could indicate a domain mismatch between the unlabeled data and the test sets.

To investigate the impact of noisy audio files, we also conduct experiments with filtered \ref{subse:cr_filtering} datasets amount of 2k and 8k hours. The results indicate that in most cases, the WER either remains unchanged or increases, suggesting that filtering does not consistently enhance model performance and sometimes even has a detrimental effect.

The model trained with 2k hours of unlabeled data is further refined in the third stage using the proposed \NemoIPL{}. The teacher model is created by averaging the top-$3$ checkpoints of the student model based on its performance on the MCV validation set. This approach results in an additional 2\% relative improvement on the MCV and Fisher test sets while maintaining stable performance on MLS and VP. Unlike simply adding more data, which leads to performance degradation on some datasets, the \NemoIPL{} approach provides a more consistent and controlled improvement, yielding an overall relative WER reduction of 2.7\% to 8\% across different benchmarks.

\begin{table}[t]
    \caption{Performance evaluation of models trained on Portugese data in terms of WER (\%). Bold values denote the best result in each group.}
    \label{tab:results_portugese}
    \vspace{-0.2cm} 
    \resizebox{\columnwidth}{!}{
    \begin{tabular}{l@{\hskip 3pt}|c@{\hskip 3pt}|c@{\hskip 3pt}c@{\hskip 3pt}c@{\hskip 3pt}c}

        \toprule
        \textbf{Model} & \textbf{Unlabeled} & \textbf{MCV} & \textbf{mTEDx} & \textbf{MLS} & \textbf{CORAA} \\
        \midrule
         Baseline CTC  & - & 15.58 & 44.15 & 33.44 & 61.65 \\
         Baseline RNNT & - & 14.85 & 39.86& 32.27 & 59.41 \\
        \midrule
        \multirow{4}{*}{\shortstack[l]{First Stage \\ No CR Filtering }}  
         & MLS & 14.5 & 39.26 & 26.32 & 56.11\\
         & 1k h & 13.72 & 29.25 & 26.71 & 47.83 \\
         & \textbf{2k h}  & \textbf{13.06} & \textbf{27.39} & \textbf{25.77} & \textbf{45.39} \\
         & 2.5k h & 13.43 & 28.33 & 27.18 & 47.72\\
         \midrule
         + CR Filtering & 2k h & 12.99 & 27.32 & \textbf{25.54} & 45.94 \\
         \midrule
         \NemoIPL{} CTC & 2k h & \textbf{12.82} & \textbf{26.7} & \textbf{25.7}& \textbf{44.84}\\
         \NemoIPL{} RNNT & 2k h & \textbf{12.03} & \textbf{25.44} & \textbf{24.78} & \textbf{42.97}\\
        \bottomrule
    \end{tabular}%
    }
\end{table}

\subsection{Portuguese language}
Table~\ref{tab:results_portugese} presents the performance evaluation of Portuguese models trained with different amounts of unlabeled data at various stages of the algorithm. With a relatively small labeled dataset (80 hours), the introduction of excessive unlabeled data (2.5k hours) led to a deterioration in accuracy. The optimal performance is observed when 2k hours of unlabeled data is added, leading to the most significant reductions in WER, particularly on the mTEDx and CORAA datasets, where absolute WER decreases by 16.76\% and 16.26\%, respectively.

We also filter the dataset to include only 2k hours of clean data and train the algorithm on this subset. Since the filtered \ref{subse:cr_filtering} data accounts for approximately 5\% of the entire dataset, the effect is not substantial. Moreover, the WER on CORAA slightly worsens, we believe this is due to the overall noisiness of the dataset; removing such audio files from the training set may have had an adverse impact.
The model trained with 2k hours of unlabeled data is further refined in the third stage using the proposed \NemoIPL{}. The teacher model is generated by averaging the top-$3$ student model checkpoints, selected based on MCV validation set performance. This refinement results in an additional 2-2.5\% relative WER improvement across most test sets, culminating in an overall relative WER reduction of 17.7\% to 39.5\% across different benchmarks.

\subsection{Ablation study}
To enable the community to assess the effectiveness of \NemoIPL{} more, we conducted a series of experiments on the LibriSpeech-100/860 benchmark, comparing it with various pseudo-labeling strategies. Our goal was to understand how different training approaches impact model performance in terms of WER.

We begin with a baseline model, trained on the 100-hour clean subset of LibriSpeech for $n_\text{epochs} =  150$ epochs. The remaining 860 hours of unlabeled speech data serve as the pool for pseudo-labeling experiments. The first step in our comparison involves generating pseudo-labels (PLs) once and using them as labeled data combined with LibriSpeech-100. This model was trained for 75 epochs, initializing from the baseline checkpoint.

Next, we explore first stage (FS) of our algorithm trained for 75 (FS\textsubscript{75}) and 150 (FS\textsubscript{150}) epochs. The former serves as the starting point for the next stage, where we introduce student-teacher learning technique. We compare the well-known EMA approach with our proposed \NemoIPL{}, both trained for an additional 75 epochs.

Table~\ref{tab:ls100-ls860} presents the results of these experiments. As expected, incorporating pseudo-labels significantly improves performance over the baseline. We observe that iterative pseudo-labeling techniques outperform single-step pseudo-labeling and benefit from extended training. While the EMA approach provides notable gains, it falls short of the improvements achieved by FS\textsubscript{150} iterative training. In contrast, despite its simplicity, \NemoIPL{} surpasses all other techniques, demonstrating its effectiveness in enhancing model performance.

\begin{table}[t]
    \centering
    \caption{Comparison of different approaches on LibriSpeech-100/LibriSpeech-860 subsets in terms of WER (\%).}
    \label{tab:ls100-ls860}
    \vspace{-0.2cm} 
    \resizebox{\columnwidth}{!}{
    \setlength\tabcolsep{1.0em}
    \begin{tabular}{llcccc}
        \toprule
        \multicolumn{2}{c}{\textbf{Model}} & \multicolumn{2}{c}{\textbf{Test-clean}} & \multicolumn{2}{c}{\textbf{Test-other}} \\
        \cmidrule(lr){3-4} \cmidrule(lr){5-6}
        & & \textbf{CTC} & \textbf{RNNT} & \textbf{CTC} & \textbf{RNNT} \\
        \midrule
        Baseline & & 15.67 & 12.69 & 35.78 & 31.01 \\
        \midrule
        PL once & & 9.16  & 8.57 & 24.23 & 22.48  \\
        FS\textsubscript{75}& & 6.12  & 5.81 & 12.15  & 11.39  \\
        FS\textsubscript{150} & & 5.56  & 5.35 & 10.49  & 9.90 \\
        \midrule
        EMA  & & 5.87  & 5.58 & 11.59 & 10.91 \\        
        
        TopIPL & & \textbf{5.30} & \textbf{5.13} & \textbf{10.07} & \textbf{9.41}   \\
        \bottomrule
    \end{tabular}%
    }
\end{table}

\section{Conclusion}
In this work, we introduce an open-source data preprocessing pipeline alongside a novel pseudo-labeling algorithm, \NemoIPL{}, for semi-supervised ASR training. We evaluate \NemoIPL{} across three distinct setups: low-resource Portuguese and Armenian, and a high-resource Spanish observing consistent relative WER reductions in all cases.  Our results show that \NemoIPL{} significantly improves model performance in low-resource scenarios, achieving up to 40\% relative WER reduction in Portuguese and 16\% in Armenian, and continues to provide meaningful gains even with abundant labeled data, with 8\% relative WER reduction in Spanish. Additionally, we show that \NemoIPL{} outperforms existing approaches and is robust to noisy unlabeled audio data. This makes it practical for real-world applications where speech data is often collected from diverse and imperfect sources. By making the entire pipeline publicly available, we establish a strong baseline for future research in semi-supervised ASR.

\bibliographystyle{IEEEtran}
\bibliography{mybib}

\end{document}